\documentclass{elsart}

\usepackage{epsfig}

\newcommand{\beq}{\begin{equation}}
\newcommand{\eeq}{\end{equation}}

\newcommand{\eq}[1]{(\ref{#1})}

\begin{document}

\begin{frontmatter}

\title{Hadronic Vacuum Polarization Contribution to the Muonium Hyperfine Splitting}

\author[vniim,mpq]{Savely G. Karshenboim}
\ead{
sek@mpq.mpg.de}
\and
\author[vniim]{Valery A. Shelyuto}
\address[vniim]{D. I. Mendeleev Institute for Metrology (VNIIM), St. Petersburg 198005, Russia}
\address[mpq]{Max-Planck-Institut f\"ur Quantenoptik, 85748 Garching, Germany}

\begin{abstract}
We discuss hadronic effects in the muonium hyperfine structure and 
derive an expression for the hadronic contribution to the {\em hfs} interval 
in form of the one-dimensional integral of the cross section of $e^+e^-$ 
annihilation into hadrons. Higher-order 
hadronic contributions are also considered.
\end{abstract}

\end{frontmatter}

\section{Introduction and Results}

\begin{figure} 
\begin{minipage}[b]{0.35\textwidth}
\centerline{\epsfbox{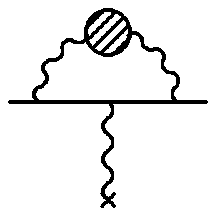}}
\end{minipage}%
\hskip 0.08\textwidth
\begin{minipage}[b]{0.55\textwidth}
\centerline{\epsfbox{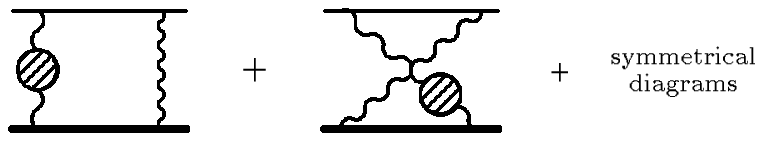}}
\end{minipage}
\vspace{10pt}
\begin{minipage}[t]{0.35\textwidth}
\caption{\label{fig1} The hadronic contribution to the anomalous magnetic moment of muon}
\end{minipage}%
\hskip 0.08\textwidth%
\begin{minipage}[t]{0.55\textwidth}
\caption{\label{fig2} The hadronic contribution to the muonium hyperfine splitting}
\end{minipage}
\end{figure}

Quantum Electrodynamics (QED) provides an opportunity to determine 
characteristics of various particles and simple atomic systems. However, any pure QED caclulation 
is incomplete even in the case of a purely leptonic system because of a 
contribution of the strong interaction originating 
from hadronic intermediate states. For example, a contribution of the hadronic 
effects to the anomalous magnetic moment of the muon ($a_\mu$) is about 60 ppm of its QED value. 
This value is larger than both  
the uncertainty of the QED calculations (see e. g. \cite{ammu2}) and experiment \cite{aexp}. 
The leading part of this correction (Fig.~\ref{fig1}) can be presented in the form 
\beq
\label{hadg}
\Delta a_\mu({\rm hadr}) = \frac{\alpha^2}{3\pi^2}\int{\frac{ds}{s}}\,K_a(s)\,R(s)
\;.
\eeq
Here $\alpha$ is the fine structure constant and relativistic units 
in which $\hbar=c=1$ will be used throughout the paper.
Such a presentation is quite useful since it clearly separates the QED and hadronic 
parts. The QED part is known \cite{ammu3} in a closed analytic form
\[
K_{\rm a} = - \biggl(\frac{s^2}{2m_\mu^4} - \frac{2s}{m_\mu^2} + 1 \biggr) \;\frac{1}{r}\,
\ln{\frac{1 + r}{1 - r}}
+ \biggl(\frac{s^2}{2m_\mu^4} - \frac{s}{m_\mu^2} \biggr) 
\ln{\frac{s}{m_\mu^2}} ~-~\frac{s}{m_\mu^2} + \frac{1}{2}\;,
\]
while the hadronic factor 
\beq
\label{Ree}
R(s) = \frac{\sigma(e^+e^-\to \gamma\to {\rm hadrons})}{4\pi\,\alpha^2/3s }
\eeq
can be determined either from direct measurements of the cross sections of $e^+e^-$-annihilation into hadrons at
the energy $E_{\rm c.m.}=\sqrt{s}$ or from theoretical estimations.
Here $r=\bigl(1-4m_\mu^2/s\bigr)^{1/2}$.

In this paper we derive a similar expression for the hyperfine structure (hfs) interval in a two-body hydrogen-like atom with 
a point-like nucleus (Fig.~\ref{fig2}). In particular, in the case of the ground state hfs in muonium the result is
\beq \label{hadmu}
\Delta E_{VP} = \frac{2}{3}\frac{\alpha^2}{\pi^2}\frac{m_e}{m_\mu}E_F
\int{\frac{ds}{s}}K_{\rm Mu}(s)R(s)\;,
\eeq
where
\beq\label{Kmu}
K_{\rm Mu}(s) = -\left(\frac{s}{4m_\mu^2}+2 \right)\;r\,
\ln{\frac{1+r}{1-r}}
+\left( \frac{s}{4m_\mu^2} +\frac{3}{2} \right)
\ln{\frac{s}{m_\mu^2}}
-\frac{1}{2}\;,
\eeq
and $E_F=8\alpha^4m_R^3/3m_e m_\mu$ is the leading contribution to the hyperfine structure (the so
called Fermi energy) and $m_R$ is the reduced mass.

Presently, the accuracy of the QED calculations \cite{icap,report,new} and of evaluations of 
the hadronic vacuum polarization contributions (see e.g. \cite{faustov}) 
is sufficient for comparison with the experiment \cite{exp}. However, the
expected progress in development of intensive muon 
sources for needs of particle physics offers an opportunity to increase the 
statistics of 
muonic events and to provide a much better source of muonium 
\cite{future}. In view of this increase of the experimental accuracy we 
need more precise theoretical 
calculations of various contributions and in particular of those for
the hadronic vacuum polarization (Fig.~\ref{fig2}). The accuracy of such
a hadronic calculation should establish the frontiers of any possible precision test of the bound state QED with muonium. 

\begin{figure}[h]
\epsfxsize=10cm
\centerline{\epsfbox{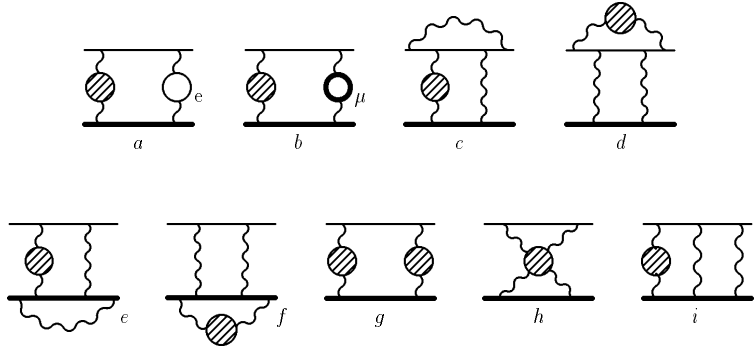}}
\caption{\label{fig5} Second-order hadronic corrections to the muonium hfs}
\end{figure}

The expression \eq{hadmu} was obtained by us some time ago and the result 
appeared in a paper of one of us \cite{zp}. 
However, neither derivation nor discussion of the corrections to 
(\ref{hadmu}) was presented. Since a calculation based on this expression is 
now in progress \cite{new},  
the derivation of \eq{hadmu} will be given here in detail. The accuracy of 
an incoming calculation \cite{new} based on our expression \eq{hadmu} 
is of the 1\%-level and we discuss here 
various corrections to \eq{hadmu} and in particular those due to the 
higher-order hadronic vacuum effects (Fig.~\ref{fig5}).
Eventually we found that there is only one higher-order correction 
to \eq{hadmu} above the uncertainty level of 
the calculation in \cite{new} and it increases the vacuum polarization 
contribution by roughly 3\%. 

\section{General expression}

We start the derivation with the two-photon exchange diagrams. Their contribution 
can be presented in the form (cf. \cite{AP})
\[
\Delta E_{Sc} = \frac{\alpha}{\pi}\frac{m_e m_\mu}{m_\mu^2-m_e^2}E_F \cdot
\Big[I(m_\mu) - I(m_e)\Big]\;,
~{\rm where}~
I(m)=\int_{0}^{\infty}{\frac{dk^2}{k^2}J(k,m)}
\]
and
\beq\label{defJ}
J(k,m)=2\left(\frac{1}{k}\sqrt{k^2+4m^2}-1\right)-\frac{1}{4m^2}\left(k\sqrt{k^2+4m^2}-k^2-2m^2\right)\;.
\eeq
When calculating the integral over the
euclidean momentum $k$ a divergency at low momentum should appear and it is necessary to 
rearrange the expressions. By introducing the vacuum 
polarization insertion in 
the skeleton integral \eq{defJ}, we reach our goal without any additional rearrangement.

We take the polarization into account substituting the photon propagator with a dispersion presentation of the polarization
\beq    \label{Imag}
\frac{1}{k^2}  \to  \frac{\alpha}{\pi}\int
{\frac{
d s~ \rho(s)}{k^2+s}}\;.
\eeq
The choice of the dispersion weight function $\rho(s)$ and the limits of the integration depend 
on what contribution is to be calculated. 
In the case of the hadronic vacuum polarization one finds
$s\geq 4m_\pi^2\gg m_e^2$,
and $\rho(s) =R(s)/3s$ 
where $R(s)$ is defined in \eq{Ree}.
Finally, we present the vacuum polarization contribution in the form
\beq
\Delta E_{VP} = 2\,\frac{\alpha^2}{\pi^2}\frac{m_e m_\mu}{m_\mu^2-m_e^2} E_F\cdot
\int{ds~ \rho(s)}
\Bigl[I_{VP}(s,m_\mu) - I_{VP}(s,m_e)\Bigr]
\;,
\eeq
where
\beq            \label{defIVP}
I_{VP}(s,m) =
\int_{0}^{\infty}{\frac{dk^2}{k^2+s} J(k,m)}\;.
\eeq

After integrating over the momentum $k$ we find
%
\[
I_{VP}(s,m) = - 2 \left[2+ \frac{s}{4 m^2} \right]
L\bigl({s}/{4m^2}\bigl) +2\left[ \frac{3}{2} + \left(\frac{s}{4m^2}\right)^2 \right]
\ln\bigl({s}/{m^2}\bigr)-\frac{1}{2}
\;,
\]
%
where
\[
L(\tau) =
\left\{
\begin{array}{c}
\frac{\sqrt{1-\tau^2}}{\tau}
\tan^{-1}\frac{\sqrt{1-\tau^2}}{\tau}\;,~~~~\tau<1\;,\\
\frac{\sqrt{\tau^2-1}}{\tau}
\frac{1}{2}\ln{\frac{\tau+\sqrt{\tau^2-1}}
{\tau-\sqrt{\tau^2-1}}}\;,~~~~\tau\geq1\;.\\
\end{array}
\right.
\]
That is the most general expression for the hadronic contribution to the hyperfine splitting in 
a hydrogen-like system with a point-like nucleus.
In the case of the hadronic contribution to the {\em hfs} interval in muonium one can take advantage of $s > 4m_\mu^2 \gg 4m_e^2$ and
neglect a contribution of $m=m_e$
\beq  \label{deltaKmu}
\Big[
\cdots
\Big]_{m=m_e}\simeq - \frac{4m_e^2}{s}\left(\frac{9}{8}\ln{\frac{s}{m_e^2}}
+\frac{15}{16}\right)\;.
\eeq
After simple transformations we arrive to the result \eq{hadmu}. 
This expression is appropriate for calculations of any vacuum polarization contributions with $s\leq 4m_\mu^2$ such as the hadronic 
or $\tau$-leptonic contributions. The neglected term \eq{deltaKmu} does not exceed $10^{-3}$ of the main contribution for 
$s\geq (2m_\pi)^2$.

\section{Asymptotics of the QED kernels $K_{\rm Mu}(s)$ at high energy}

Since $s\geq (2m_\pi)^2 > (2m_\mu)^2$ the asymptotic behaviours of the QED kernel $K_{\rm Mu}$ at high values of $s$ is of interest.
We find
\beq \label{AsKmu}
K_{\rm Mu} = 
\frac{4m_\mu^2}{s}\left[\frac{9}{8}\ln{\frac{s}{m_\mu^2}}
+\frac{15}{16}\right]
+ \left(\frac{4m_\mu^2}{s}\right)^2\left[\frac{5}{16}\ln{\frac{s}{m_\mu^2}}
-\frac{17}{96}\right]+ \dots
\eeq
In the case of the $\rho$-meson contribution, which is dominant (see Sect.~\ref{Snar}), the parameter of expansion is $4m_\mu^2/m_\rho^2\simeq 0.076$
and the leading contribution at $s=m_\rho^2$ is $0.409$, 
while the higher order terms are only $0.006$.

\begin{figure} %
\begin{minipage}[b]{0.43\textwidth}
\epsfxsize=7.3cm
\centerline{\epsfbox{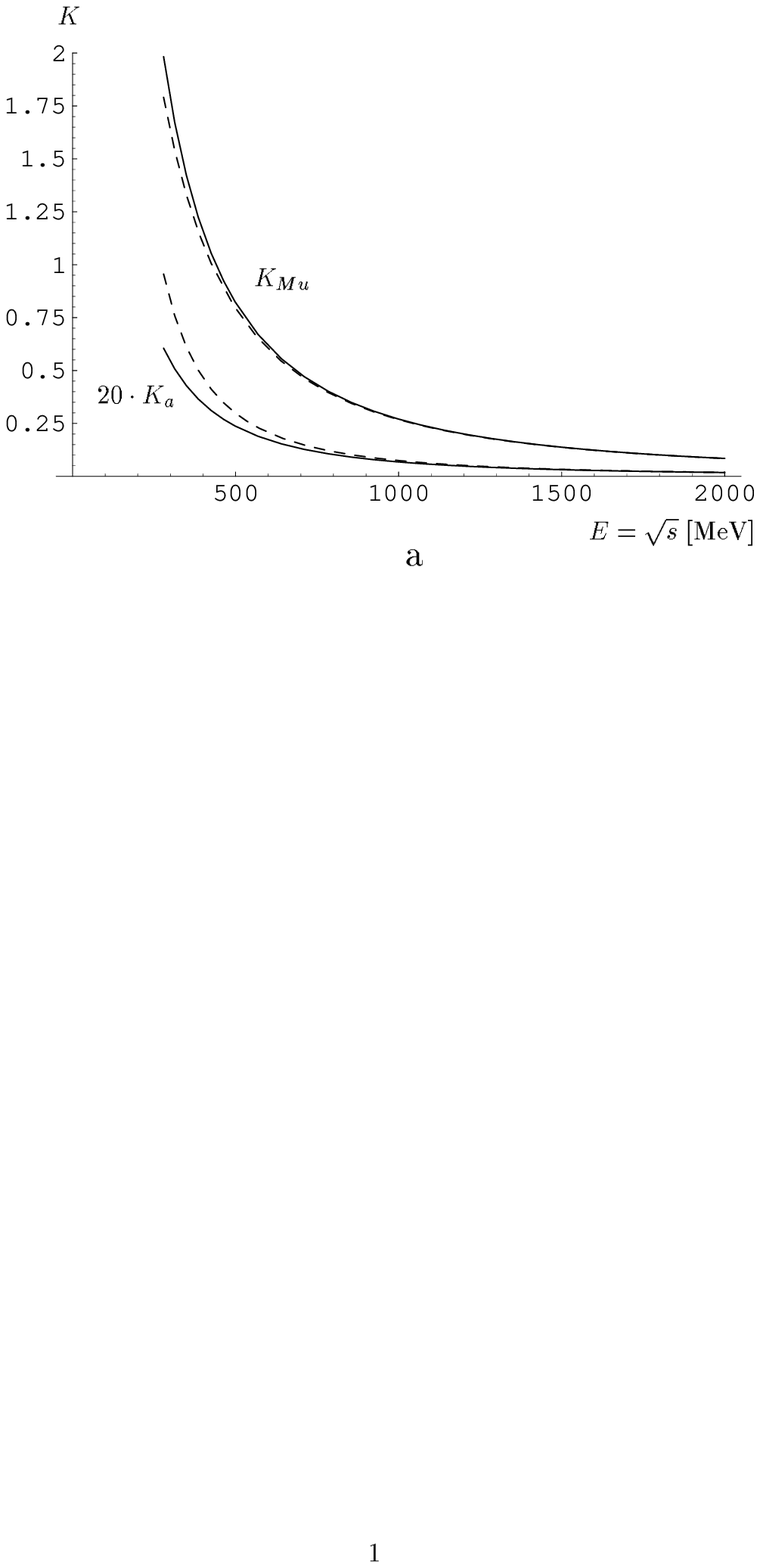}}
\end{minipage}%
\hskip 0.08\textwidth
\begin{minipage}[b]{0.43\textwidth}
\epsfxsize=7.3cm
\centerline{\epsfbox{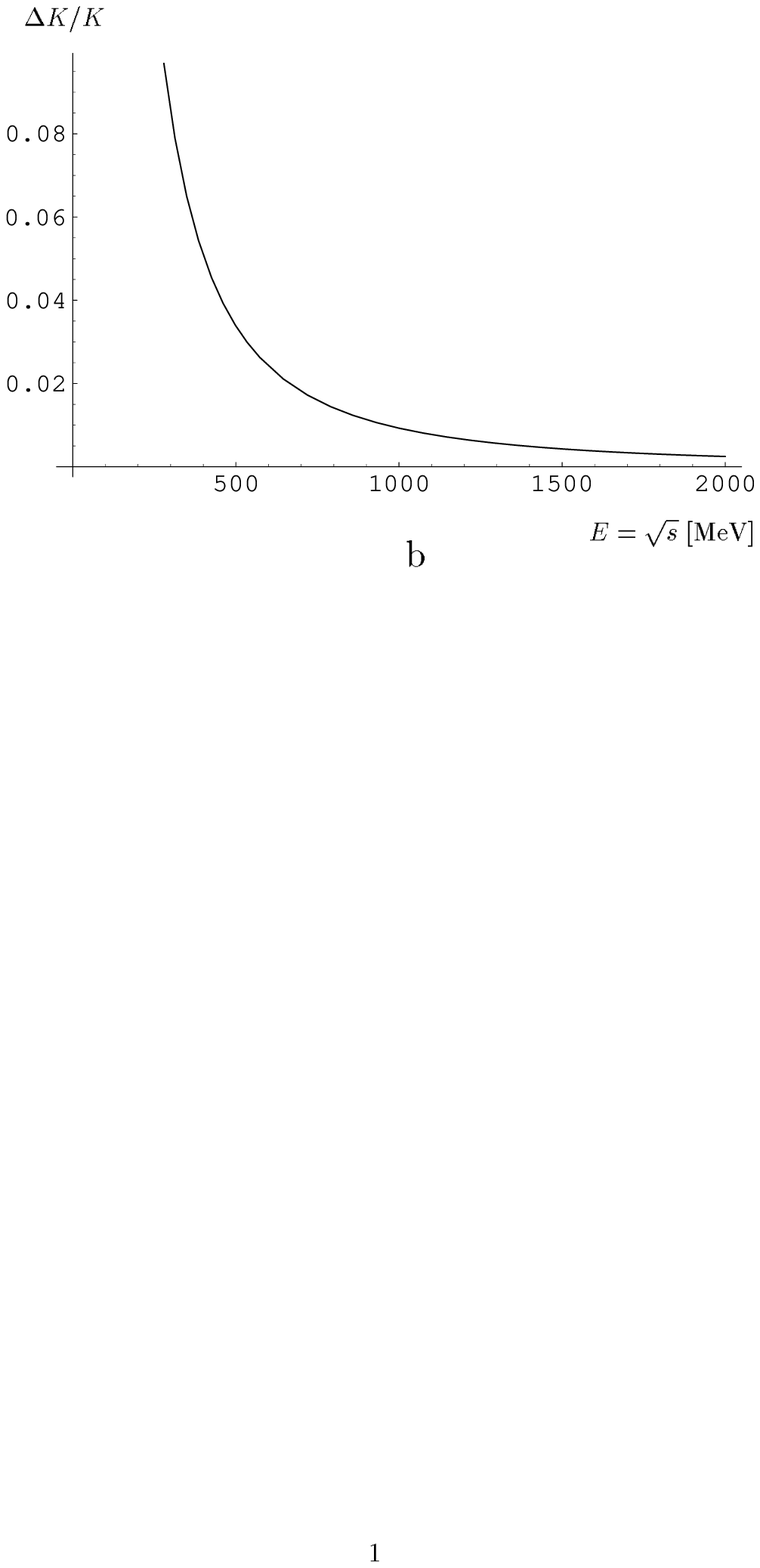}}
\end{minipage}
\vspace{10pt}
\caption{\label{fig6} The QED kernels for the hadronic contributions
to the muonium hyperfine splitting ($K_{\rm Mu}$)
and to the anomalous magnetic moment of muon ($K_a$)
and their leading asymptotics, $K_{\rm Mu}^{(0)}=\bigl(18\ln(s/m_\mu^2)+15\bigr)m_\mu^2/4s$ and 
$K_a^{(0)}=m_\mu^2/3s$. The kernels are shown in Fig.~{\em a} with solid lines, while the asymptotics are 
with dashed lines. In Fig.~{\em b} a result for 
$\bigl(K_{\rm Mu}-K_{\rm Mu}^{(0)}\bigr)/K_{\rm Mu}$ is presented.}
\end{figure}

This behaviour should be compared with the expansion of $K_a(s)$ (see Fig.~\ref{fig6})
\beq\label{Kas}
K_a(s) \simeq  \frac{1}{12}\frac{4m_\mu^2}{s} - \left(\frac{4m_\mu^2}{s}\right)^2\left[\frac{1}{16}\ln{\frac{s}{m_\mu^2}}
-\frac{25}{192}\right]+\dots
\eeq 

\section{Estimation within a simplified narrow-pole model \label{Snar}}

The essential part of the hadronic vacuum polarization contribution comes from the $\rho$-meson-pole and 
it can be easily estimated  (cf. \cite{sty}) from a simple approximation 
\beq\label{simple}
\rho_{\rm pole}(s)=\sum_{res}{}\frac{4\pi^2}{f_{\rm res}^2}\delta(s-M_{res}^2)
=\sum_{res}{}\frac{\Gamma({\rm res}\to e^+e^-)}{\alpha^2/3\pi\;M_{\rm res}}\delta(s-M_{res}^2)\;,
\eeq
with a sum over three light vector mesons dominating in $e^+e^-$ annihilation
at low energies: $\rho$, $\omega$ and $\phi$. 
The partial width of a resonance decay into 
an electron-positron pair is a value determined experimentally. 
The results for the $\rho$-, $\omega$- and $\phi$-mesons within the simple pole approximation are summarized 
in Table 1 as well as their masses and electronic widths. This calculation is important to check if our expression 
is consistent with previous calculations of the hadronic contributions. The $\rho$-meson contribution is 65\% of 
the hadronic contribution \cite{new} and all three pole contributions deliver 76\% of $\Delta \nu_{\rm hadr}$ 
(or 0.18 kHz or $4.0\cdot 10^{-8}\nu_F$). The value of the $\rho$-meson 
contribution is in a good agreement with the $\pi\pi$ contribution 
in \cite{new}. Due to that we study any correction to \eq{hadmu} with 
the help of these resonance contributions obtained from the simple pole approximation.

\begin{table}[th]
\begin{center}
\begin{tabular}{|c|c|c|c|}
\hline
Resonance & $\rho$-meson & $\omega$-meson & $\phi$-meson \\
\hline
$M_{res}$ [MeV]                                      & 769(1)             &  783                &  1019               \\
$\Gamma({\rm res}\to e^+e^-)$ [keV]                  & 6.77(32)             &  0.60(2)             &  1.32(5)            \\
$\Delta \nu_{\rm res}$ [kHz]                         &  0.151(7)              &  0.013               &  0.014          \\
${\Delta \nu_{\rm res}}/{\Delta \nu_{\rm hadr}}$ &  65(3)\%           &
  5.5(2)\%                &  6.0(2)\%            \\
$K_{\rm Mu}$                                         &  0.415           & 0.404             & 0.262              \\
$K_{\rm Mu}^{(0)}$                                   &  0.409           & 0.398             & 0.260     \\
\hline
\end{tabular}
\vspace{5mm} 
\caption{Vector meson resonances, their parameters and contributions to the hyperfine structure.
The fractional value in respect the
to the complete hadronic contribution is given for 
$\Delta \nu_{\rm hadr}=0.233(3)$ kHz \protect\cite{new}. The parameters 
are taken from \protect\cite{data}}
\end{center}
\end{table}

\section{Higher-Order Hadronic Contributions}

To estimate the higher-order hadronic contribution in order $\alpha^2(Z \alpha)$ we simply note that only one set of diagrams leads to 
a contribition enhanced by a logarithmic factor $\ln(s/m_e^2)$ 
(Fig.~\ref{fig5}a). These diagrams contain the electronic vacuum 
polarization insertion and in the logarithmic approximation can be easily calculated for the $\rho$-meson contribution:
\beq\label{lolog}
\Delta E ({\rm log}) = 
3\cdot\frac{\alpha}{3\pi}\cdot\ln\frac{m_\rho^2}{(2m_e)^2}\cdot\Delta E ({\rm leading~contribution})\;.
\eeq
The same estimation can be reproduced for two other resonances in Table 1. 
About 80\% of the hadronic contribution comes from these three 
resonances, while the rest is essentially related to the high-$s$ backdround. 
It it easy to check that in the case of the 
background contribution and other resonances the logarithmic approximation is valid and we confirm the logarithmic approximation
(cf. (\ref{lolog})) for the complete result, 
which is about 3\% of the leading contribution. We expect that a non-logarithmic part of the cotributions (Fig.~\ref{fig5}) is below 1\%-level. 

A similar estimation is misleading in the case of the anomalous 
magnetic moment of the muon \cite{amuh}, where the contribution related to insertion 
of the electronic vacuum polarization into leading hadronic diagram Fig. 1 does not dominate among higher-order 
hadronic contributions. We believe that it is caused by a special 
structure of the kernel (\ref{Kas}), which has a small numerical 
coefficient and no logarithmic enhancement at high $s$. 
That is not a case for the muonium hfs (\ref{AsKmu}) and we finally 
estimate the higher order hadronic contribution as $(3\pm1)\%$ of the 
leading hadronic term, or 0.007(2) kHz.

\subsection*{Acknowledgements}

We are grateful to A. Czarmecki, M. Eides, K. Jungmann and espesially to S. Eidelman for stimulating discussions.  The
work was supported in part by the Russian Foundation for
Basic Researches (under grant \# 00-02-16718) and the Russian State
Program ``Fundamental Metrology''.

\end{document}